# Your Computer is Leaking

Ian J. Malloy

Dennis Hollenbeck, 149 W Managed Services

dennis.hollenbeck@149w-managed-services.com

**Abstract**

*This presentation focuses on differences between quantum computing and quantum cryptography. Both are discussed related to classical computer systems in terms of vulnerability. Research concerning quantum cryptography is analyzed in terms of work done by the University of Cambridge in partnership with a division of Toshiba, and also attacks demonstrated by Swedish researchers against QKD of energy-time entangled systems. Quantum computing is covered in terms of classical cryptography related to weaknesses presented by Shor's algorithm. Previous classical vulnerabilities also discussed were conducted by Israeli researchers as a side-channel attack using parabolic curve microphones, which has since been patched.*

**Keywords**

Post-quantum cryptography, encryption, quantum computing

## INTRODUCTION

Before beginning, it's useful to introduce the fictional characters Alice, Bob, and Oscar. Alice and Bob are used in cryptography to make concepts easier to understand. Oscar is traditionally used as an adversary while Alice and Bob work to communicate securely. These characters will be referenced several times throughout this presentation. To set the background of this presentation, work done by researchers at Tel-Aviv University proved acoustic signals allowed side-channel attacks against several target computers [1]. Side-channel attacks are a form of cryptanalysis, or way to break encryption, that focuses on physical components as opposed to algorithmic weaknesses. Not only was it possible to conduct this attack, it was also possible to increase the distance the attack could be performed over using a parabolic reflector to amplify the signal.

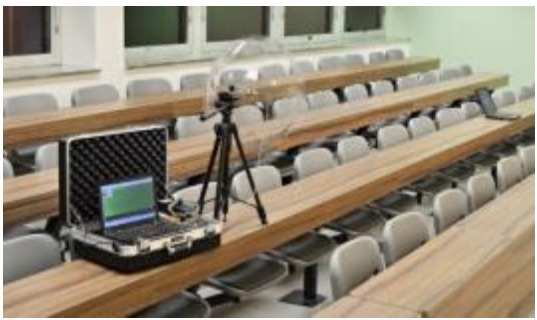

*Figure 1.* RSA GNuPG Side-Channel Microphone

There was a proclaimed patch issued that fixed this vulnerability, but there are always ways to improve attacks. This research is important to keep in mind because it demonstrates both creative ways to crack encryption and also how radio waves are potentially tied to acoustic signals generated by data processes. This specific attack covered a wide-range of devices but focused on a specific implementation of a specific algorithm. Regardless, the risks are real and have a potential to be increased significantly. Shifting from classical cryptography and communications, there is a strong focus on developing quantum computing and quantum cryptography. Quantum computing is based on the qubit, geometrically represented by a Bloch Sphere, and quantum cryptography which involves quantum key distribution (QKD) to secure communications between Alice and Bob.

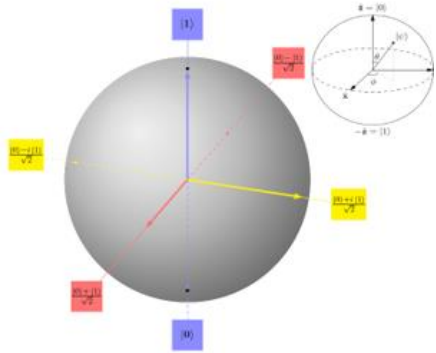

*Figure 2*. Bloch Sphere Representations

**RELEVANT LITERATURE**

The University of Cambridge partnered with a research division of Toshiba and were able to use a combination of "lit" and "dark" telecommunication fibers to effectively use QKD to secure communications [2]. Quantum key distribution is equivalent to a one-time pad but doesn't require Alice and Bob to physically meet to share keys beforehand. It was assumed that QKD was essentially bullet proof, or that it couldn't be hacked, since any attempt by Oscar to listen in on Alice and Bob would alert Alice and Bob immediately. Basically Alice and Bob are observing their communications so that if Oscar tried anything it would change measurements that Alice and Bob are conducting on particles. A demonstration of how Oscar can listen to Alice and Bob without triggering changes in measurement of the particles was conducted by researchers towards the end of 2015 [3].

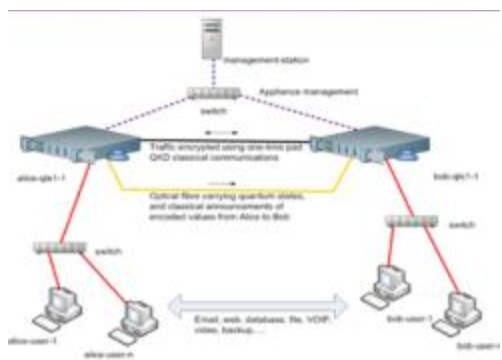

*Figure 3*. Quantum Key Distribution – Cambridge-Toshiba Experiment

Essentially, it was shown that Oscar could hack Bell's inequality using classical light sources to spoof the measurements Alice and Bob expected. Bell's inequality is an expected statistical result that occurs after particles are entangled and then separated. In Quantum Key Distribution, the difference from a one time pad is that QKD doesn't require a rendezvous between Alice and Bob to exchange keys. The expectation is that once Alice and Bob begin communicating, any attempts by Oscar will spoil the measurements of Alice and Bob.

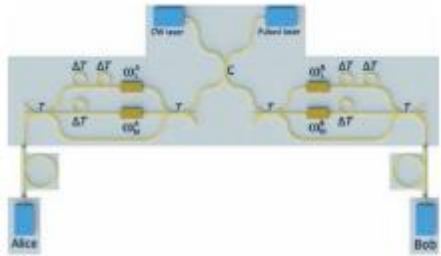

*Figure 4*. Quantum Key Distribution – Swedish Bell Hack

Alice and Bob rely on their measurements during communication to change if Oscar can tap into their conversation. If a Quantum Key Distribution system can be tapped into by using classical optical devices to manipulate the sensors used by Alice and Bob during communication, Oscar wins. If Alice and Bob trust that any interference by Oscar necessarily changes their measurements, it can be shown that Oscar can still tap into the device used by researchers at the University of Sweden in Stockholm for purposes of demonstration. Thankfully the researchers who demonstrated the ability to hack this form of QKD also proposed solutions to the problem [3].

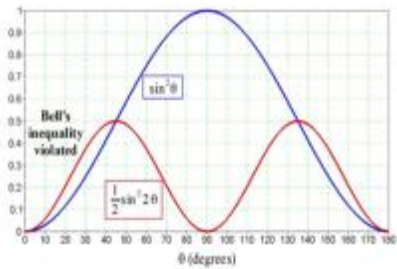

The United States National Security Agency (NSA) and National Institute of Standards and Technology (NIST) have moved forward with requests for input on post-quantum cryptography [4]. Quantum computing was first conceptualized as a form of Turing machine, but a scientist named Peter Shor developed an algorithm, now known as Shor's Algorithm which advanced concepts of quantum computing by providing a verifiable algorithm a quantum computer could run. Shor's algorithm solves factorization problems very efficiently.

**VULNERABILITIES IN ENCRYPTION**

The weaknesses of discrete logarithm problems and cryptography based upon integer factorization problems become evident when you start to understand what a qubit is able to do computationally. If you think of a classical bit, some form of 0's or 1's, and then move to a qubit that can be both 0 and 1 the ability to combine calculations is significantly advanced. If you can keep several qubits in a state of coherence, and each can enter states of superposition to approximate solutions to problems, then breaking certain forms of encryption is simplified.

Disclosures by Edward Snowden revealed an interest of the NSA to actively research the feasibility and requirements to build a quantum computer specifically designed to break encryption [5]. The NSA has also released a notice of adjustments to their Suite-B cryptography which first appeared as a desire to seek post-quantum cryptography in the "not too distant future," which appears to be now given the move for NIST requesting expert advice. NIST is interested in proposals for new standards in encryption that will survive in a post-quantum world.

## CHALLENGES IN QUANTUM COMPUTING

Vulnerable cryptography in a post-quantum world shares the same threat in common, Shor's algorithm. Any system now or in the future that relies on either integer factorization or discrete logarithm problems will be vulnerable to cryptanalysis from quantum computers implementing Shor's algorithm. There are still issues with controlling enough qubits to factor numbers as large as those used in modern cryptography, and this is a critical challenge for researchers in quantum computing.

While the NSA has been working on establishing the feasibility of a quantum computer to break encryption, whether they succeed or not still raises the question of how they would deploy this type of system. With the move from seeking "in the near future" to the current request released by NIST concerning post-quantum cryptography, anticipating quantum computing advances is important. There are a lot of fast-paced discoveries being made by several teams in quantum computing research, and it's a good idea to begin taking steps to address this now.

## DISCUSSION

There are too many networks with open vulnerabilities that have had the patches available but weren't updated to address these same vulnerabilities. Most security professionals have experienced this directly. Stories released by the SANS Institute in their "Newsbites" have editorial comments, and many involve these types of issues. Quantum mechanics is really weird science, the word "quantum" itself can complicate any discussion that follows. One of the hardest components of cyber security hinges on educating people about why they need to invest in security, and educating these same people on what they can do to address the problem.

If a company doesn't think they could be targeted, or think they haven't been targeted it's easy to imagine how hard it would be to explain this new threat. A CEO might not easily believe that networks could be threatened by a computer that works with just a few a particles of light. This is an exaggeration, but also a threat the NSA takes seriously. Then again, companies might think their data doesn't need the same level of security that a government agency would recommend.

By working with a Riemann-Hilbert intersection and developing algorithms, it was possible to derive a potential way to mitigate quantum based attacks [6]. By relying on radio waves, which are a form of electromagnetism as well as a form of light, it seems plausible enough to research further. The Riemann sphere transitions over time from spherical to a single point, leading to a complex algorithm producing several vectors perpendicular to the surface, and repeating the process. While this is a preliminary finding, it remains promising.

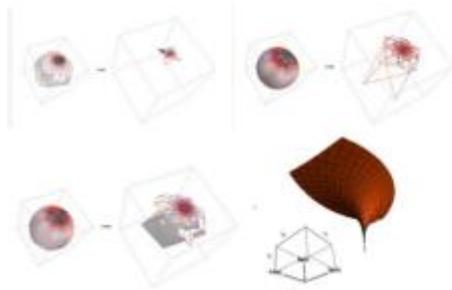

*Figure 5*. Riemann-Hilbert Intersection

The threats to encryption posed by Shor's algorithm dominate most news articles aimed at general audiences, but there are important things to keep in mind. First, news articles about science aimed at the general public tend to water down the science. Second, Shor's algorithm requires several qubits to maintain coherence long enough to factor numbers and this is a difficulty in quantum computing. A researcher at the University of New South Wales explained quantum computing using the analogy of the discovery of the laser. The scientist responsible for the laser didn't envision it being used to read data from a c.d., but oftentimes the scientific community advances discoveries through novel applications.

**CONCLUSION**

A discovery doesn't grant omniscience in terms of its applications, you have to hold it in your hand first to start to see how it could be used. Computation might seem different, but security isn't and you can't hold cyber security in your hand. If the threats today advance as dynamically as they are known too, when quantum computing transitions fully from exploration to application we won't know all the changes that will occur across a threat-landscape. Even though security professionals don't have complete foresight, it doesn't mean the threat shouldn't be anticipated in some form.